\documentclass[twocolumn,showpacs,preprintnumbers,amsmath,amssymb]{revtex4}
\preprint{}
\usepackage{graphicx}% Include figure files
\usepackage{dcolumn}% Align table columns on decimal point
\usepackage{bm}% bold math
\usepackage{mathrsfs}
\begin{document}
\title{Generation of multi-photon entanglement}
\author{Dong  Xie}
\email{xiedong@mail.ustc.edu.cn}
\author{An Min Wang}
 \email{anmwang@ustc.edu.cn}
  \affiliation{Department of Modern Physics , University of Science and Technology of China, Hefei, Anhui, China.}
\begin{abstract}
We propose a new scheme to generate the multi-photon entanglement via two
steps, that is, first to utilize the superconductor to create the
multi-quantum-dot entanglement, and then to use the input photon to transfer
it into the multi-photon entanglement. Moreover, the maximum probability for
the swap of photon and quantum-dot qubits is close to unit for a single
input Gaussian photon. More importantly, by mapping the multi-quantum-dot
state into the coherent states of oscillators, such as cavity modes, the
multi-quantum-dot entanglement in our scheme can be protected from the
decoherence induced by the noise. Thus, it is possible to generate more than
eight spatially separated entangled photons in the realistic experimental
conditions.
\end{abstract}
\maketitle
\section{Introduction}
Creation of large multipartite entangled states is a fundamental scientific interest \cite{lab1} and the enabling technology for quantum information and quantum networking \cite{lab2}. Due to the large spread speed of photon, multi-photon entanglement attracts widespread attention.
Up to now, the most widely used method of generating multi-photon entanglement is that prepare several pairs entangled photons based on the spontaneous parametric down-conversion in a nonlinear crystal \cite{lab3}, and use interferometer to combine them into the multi-photon entangled state, such as the Schr$\ddot{\textmd{o}}$dinger cat state. Via this process, the probability of generating multi-photon entanglement reduces exponentially with the increasing of photon number.  Experimentally, six spatially separated single photons were entangled based on parametric down-conversion by different groups \cite{lab4,lab5,lab6,lab7,lab8}.  So far, up to eight spatially separated single photons have been created in the experiment \cite{lab9,lab10}. However, in trapped ions, 14-qubit Greenberger-Horne-Zeilinger (GHZ) states have been generated \cite{lab11}. So it is possible to create more than eight entangled photons by transferring the multi-particle entanglement to multi-photon entanglement.

We consider that use multi-quantum-dot entanglement to create the multi-photon entanglement, due to single photon created by quantum dot more easily than other system \cite{lab12}. And, by the superconductor, one can build strong long-range interaction between two quantum dots \cite{lab13}. Due to the strong interaction, it has advantage on forming multi-photon entanglement (e.g. spending a few time).

With the increasing of the number of quantum dots, the decoherence induced by noise becomes more and more remarkable. It is the main barrier for creating the multi-quantum-dot entanglement. For protecting the coherence, temporally map the multi-quantum-dot state into the coherent states of harmonic oscillators. The harmonic oscillator can be a cavity mode. The dominant decoherence channel in a cavity is photon damping. Zaki Leghtas \textit{et al.} have proposed a protection protocol for a single cavity mode, which
significantly reduces the error rate due to photon loss \cite{lab14}. We generalize the error correction protocol from a single cavity mode to many cavity modes.  This will help to create more than 14 qubits GHZ state.

Finally, input photons to swap the multi-photon entanglement with the multi-quantum-dot entanglement. Recently, direct mapping of the spin quantum state of $\textmd{Ce}^{3+}$ ion onto the
polarization state of the emitted photon was completed, and the degree of spin polarization was higher than $99\%$ \cite{lab15}. Those experiment establishes the foundation for creating multi-photon entanglement.

The rest of this article is arranged as follows. In the sections II, the multi-quantum-dot entanglement is generated by superconductors. In the section III, utilize the cavity modes to protect the multi-quantum-dot entanglement state. The multi-photon entanglement is created by the input photons in the section IV. We draw up the conclusion in the section V.
\section{multi-quantum-dot entanglement}
To form the multi-quantum-dot entanglement, we consider following two kinds of interaction Hamiltonian between quantum dot and superconductor.

The first kind: two quantum dots $(a, b)$, each of which contains one electron  and is connected to two superconducting leads $(L, R)$ by tunnel junctions \cite{lab16}, as shown in Fig. 1. The effective interaction Hamiltonian is given by ($\hbar=c=1$):
\begin{eqnarray}
H_{int}^{1}=J_1\vec{\sigma}_a\cdot\vec{\sigma}_b,
\end{eqnarray}
where $J_1$ is the coupling constant, and $\vec{\sigma}_a$ is the Pauli operator of electronic spin in the quantum dot $a$ .
\begin{figure}[h]
\includegraphics[scale=0.3]{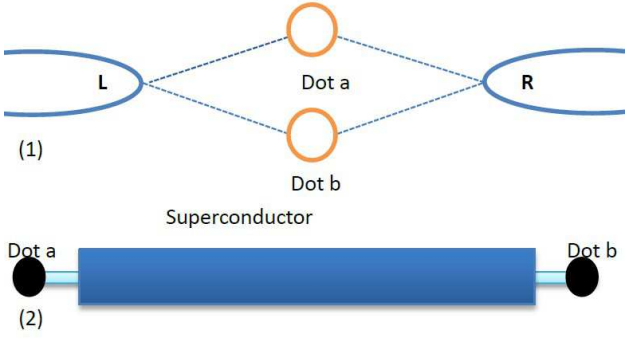}
 \caption{\label{fig.1}(1) two quantum dots are connected to two superconducting leads respectively, and the dashed lines represent the tunnel junctions; (2) two quantum dots are connected by a superconductor.  }
 \end{figure}

The second kind: two quantum dots $(a, b)$ are connected by a superconductor \cite{lab13}, as shown in Fig. 1. Via crossed Andreev reflection, two quantum dots form a strong long-range interaction. The effective interaction Hamiltonian is described by
\begin{equation}
H_{int}^{2}=J_2 \sigma_{a}^Z\otimes \sigma_{b}^Z.
\end{equation}

It deserves to note that the
quantum dots are defined in a semiconducting nanowire (e.g., InSb or InAs), which is an experimentally potential system for spin qubits \cite{lab17,lab18} and can form good interfaces with a superconductor \cite{lab19,lab20}.

Use the second interaction to create two quantum dots entanglement. Let the initial state of quantum dots to be $\displaystyle\frac{(|0\rangle+|1\rangle)^{\bigotimes2}}{2}$. Control the interaction time (for example, move away the superconductor) to be $\displaystyle t=\frac{\pi}{4J_2}$. The two quantum dots are in the maximum entangled state
\begin{equation}
\begin{split}
|\Psi\rangle=&\frac{1}{2}|0\rangle(\exp[-i\pi/4]|0\rangle+\exp[i\pi/4]|1\rangle)+\\
&\frac{1}{2}|1\rangle(\exp[i\pi/4]|0\rangle+\exp[-i\pi/4]|1\rangle).
\end{split}
\end{equation}
Then, use a $\pi/2$ pulse to act on one quantum dot. As a result, the above state is mapped into $\displaystyle\frac{|00\rangle+|11\rangle}{\sqrt{2}}$.

In order to create four quantum dots entanglement, it just need to combine two maximally entangled states of two quantum dots $\left(\displaystyle\frac{|0\rangle_1|0\rangle_2+|1\rangle_1|1\rangle_2}{\sqrt{2}}, \displaystyle\frac{|0\rangle_3|0\rangle_4+|1\rangle_3|1\rangle_4}{\sqrt{2}}\right)$ via the second interaction Hamiltonian $H_{int}^{2}$ between quantum dot 1 and quantum dot 3.

Subsequently, we can obtain the maximum entanglement state
\begin{equation}
\begin{split}
|\Psi\rangle_{1234}=\frac{1}{2}(|0\rangle|0\rangle(|0\rangle|0\rangle+|1\rangle|1\rangle)
+|1\rangle|1\rangle(|0\rangle|0\rangle-|1\rangle|1\rangle)).
\end{split}
\end{equation}
For arriving at the GHZ state, firstly, use the $\pi$ pulse to get the state
\begin{equation}
\begin{split}
|\Psi'\rangle_{1234}=&\frac{1}{2}(|0\rangle|0\rangle(|0\rangle|1\rangle+\exp[i\Delta]|1\rangle|0\rangle)\\
&+|1\rangle|1\rangle(|0\rangle|1\rangle-\exp[i\Delta]|1\rangle|0\rangle)).
\end{split}
\end{equation}
 Secondly, utilize the first interaction Hamiltonian $H_{int}^{1}$ between two quantum dots 3 and 4. Finally, we get the four qubits GHZ state. The corresponding interaction time during the first interaction Hamiltonian is given by
$\displaystyle t=\frac{\pi}{8J_1}$; and $\displaystyle\Delta=\frac{\pi}{2}$, which is the necessary condition for generating the GHZ state.

To get the GHZ state for six quantum dots, as the above way combine four quantum dots entanglement and two quantum dots
entanglement. By that analogy, one can get the multi-quantum-dots entanglement by just controlling the interaction time, not requiring the complex technique. So it is operative in the experiment.

The total time that is required to create $n$ quantum dots entanglement is given by (omitting the time of pulse)
\begin{equation}
\textmd{T}_{\textmd{total}}=[\frac{n+1}{2}]\frac{\pi}{4J_2}+[\frac{n-1}{2}]\frac{\pi}{8J_1},
\end{equation}
where we define that $\displaystyle[\frac{n}{2}]=n/2$ when $n$ is even, otherwise $\displaystyle[\frac{n}{2}]=(n-1)/2$. For strong coupling, $J_1$ and $J_2$ $\approx$ $10^8$ Hz \cite{lab13}, the total time $\displaystyle\textmd{T}_{\textmd{total}}\approx n*10^{-8}$s.
\section{Protection of the coherence}
With the increasing of entangled particles, the decoherence effects become more and more strong due to new decoherence channels.  The harmonic oscillators can be chosen to protect the coherence, because it is an infinite system and provides a vast Hilbert space without adding new dechoerence channels. For the cavity mode, the main decoherence channel comes from the photon damping. It can be corrected by a quantum non demolition parity measurement \cite{lab14}.

For protecting the coherence, it is necessary to create the entangled state between cavity modes and quantum dots by using the unitary operation,
\begin{equation}
\begin{split}
\textrm{U}_{\textmd{encode}}(|11\rangle+|00\rangle)\otimes|0\rangle_{\textmd{cavity}}=&|0\rangle\otimes(|0\mathcal{C}_\alpha^+\rangle+|1\mathcal{C}_{i\alpha}^+\rangle)\\
&+O(e^{-|\alpha|^2}),
\end{split}
\end{equation}
where $|\mathcal{C}_\alpha^\pm\rangle=\mathcal{N}(|\alpha\rangle\pm|-\alpha\rangle)$, and $|\mathcal{C}_{i\alpha}^\pm\rangle=\mathcal{N}(|i\alpha\rangle\pm|-i\alpha\rangle)$. The normalizing factor $\mathcal{N}\approx1/\sqrt{2}$, and $|\alpha\rangle$ denotes a coherent state of complex amplitude $\alpha$.
The error term $O(e^{-|\alpha|^2})$ is generated by the fact that the two states $|\mathcal{C}_\alpha^+\rangle$ and $|\mathcal{C}_{i\alpha}^+\rangle$ are not exactly orthogonal. For large photon number $|\alpha|^2$, the error term is approximate to $0$.
Next, utilizing two kinds of interaction Hamiltonian as the way in the above section to combine the state $\displaystyle\frac{|0\mathcal{C}_\alpha^+\rangle+|1\mathcal{C}_{i\alpha}^+\rangle}{\sqrt{2}}$ and $\displaystyle\frac{|00\rangle+|11\rangle}{\sqrt{2}}$ into the entangled state $\displaystyle\frac{|000\mathcal{C}_\alpha^+\rangle+|111\mathcal{C}_{i\alpha}^+\rangle}{\sqrt{2}}$. Use the unitary operations to arrive at $\displaystyle\frac{|0\mathcal{C}_\alpha^+\mathcal{C}_\alpha^+\mathcal{C}_\alpha^+\rangle+|1\mathcal{C}_{i\alpha}^+\mathcal{C}_{i\alpha}^+\mathcal{C}_{i\alpha}^+\rangle}{\sqrt{2}}$.
In this way, the decoherence channels mainly come from the cavity modes.
\begin{figure}[h]
\includegraphics[scale=0.30]{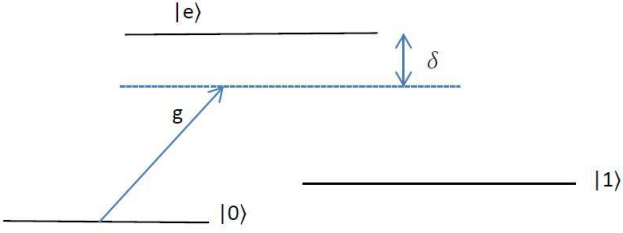}
 \caption{\label{fig.2} The detuned coupling of the state $|0\rangle$ to the excite state $|e\rangle$. Here, $g$ is the coupling constant and $\delta$ is the detuning.  }
 \end{figure}

The unitary operations are proposed in the ref. \cite{lab14,lab21}. The dispersive cavity-qubit coupling is necessary. For a quantum dot, the dispersive coupling can be offered by the detuning $\delta$, as shown in Fig. 2. The effective Hamiltonian is described by \cite{lab22}
\begin{equation}
H_d=\frac{g^2}{\delta}|0\rangle\langle0|a^\dagger a
\end{equation}
The correcting error protocol for many cavity modes can be generalized from the single cavity mode in the ref. \cite{lab14}. Perform the quantum non demolition parity measurements (the parity operator is defined by $\exp(i\pi a^\dagger a)$) independently in every cavity. If the result of parity measurement
\begin{equation}
\begin{split}
 &\left(\frac{\langle0|\langle\mathcal{C}_\alpha^+|^{\otimes n}+\langle1|\langle\mathcal{C}_{i\alpha}^+|^{\otimes n}}{\sqrt{2}}\right)\exp(i\pi a_j^\dagger a_j)\\
 &\left(\frac{|0\rangle|\mathcal{C}_\alpha^+\rangle^{\otimes n}+|1\rangle|\mathcal{C}_{i\alpha}^+\rangle^{\otimes n}}{\sqrt{2}}\right)=-1,\\
 &j\in\{1, ..., n\}
 \end{split}
\end{equation}
 it means a quantum jump in the $j$th cavity due to a photon loss. Then, repump the decayed energy
back into the cavity. For the many jumps, it is similar with the method in the ref. \cite{lab14}.
\section{multi-photon entanglement}
Now, transfer the entangled state of many cavity modes to the entangled state of multi-quantum-dot for creating the multi-photon entanglement. In another word, look for an unitary operator
\begin{equation}
U_t(|0\rangle^{\otimes n}+|1\rangle^{\otimes n})|0\rangle^{\otimes n}_{\textmd{ph}}=|0\rangle^{\otimes n}(|0\rangle^{\otimes n}+|1\rangle^{\otimes n})_{\textmd{ph}},
\end{equation}
here, $|i\rangle_{\textmd{ph}}$ denotes the state vector of photon (the photon number $i=0,1$).

\begin{figure}[h]
\includegraphics[scale=0.3]{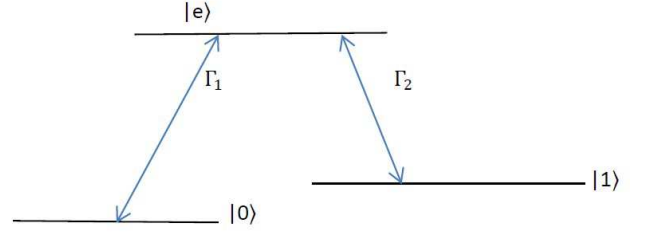}
 \caption{\label{fig.3} The tree-level quantum dot system can realize the swap between quantum dot state and photon state. $\Gamma_1$ and $\Gamma_2$ are the decay rates for the transitions from the excited state $|e\rangle$ to states $|0\rangle$ and $|1\rangle$.  }
 \end{figure}

For the single quantum dot system, the Hamiltonian is given by \cite{lab23}
\begin{equation}
\begin{split}
\mathcal{ H}=&w_1|e\rangle\langle e|+w_2|1\rangle\langle1|+\int dk(ka_k^\dagger a_k\\
&+[i(\sqrt{\frac{\Gamma_1}{2\pi}}|e\rangle\langle 0|a_k+\sqrt{\frac{\Gamma_2}{2\pi}}|e\rangle\langle 1|a_k)+H.c.]),
\end{split}
\end{equation}
where $w_1$ and $w_2$ represents the energy levels of state $|0\rangle$ and $|1\rangle$.
Excite the state  $|0\rangle$ to the state $|e\rangle$, then decay into the state $|1\rangle$. So if the initial state is the state $|0\rangle$, one can generate a photon with the frequency $(w_1-w_2)$; if the initial state
is the state $|1\rangle$, system remains unchanged . This process realizes the swap: $\displaystyle(|0\rangle+|1\rangle )|0\rangle_{\textmd{ph}}\longrightarrow |1\rangle(|0\rangle+|1\rangle )_{\textmd{ph}}$.

Input a single photon to excite the state $|0\rangle$. The initial state vector of input photon is given by
\begin{equation}
|\Phi\rangle_{\textmd{in}}=\int dkf(k)a_k^\dagger|0\rangle.
\end{equation}
The output state can be written as
\begin{equation}
\begin{split}
|\Phi\rangle_{\textmd{out}}=&\int dk(g_1(k)a_k^\dagger|0\rangle_{\textmd{ph}}|0\rangle+g_2(k)a_k^\dagger|0\rangle_{\textmd{ph}}|1\rangle\\
&+g_3|0\rangle_{\textmd{ph}}|e\rangle.
\end{split}
\end{equation}
Utilize the Schr$\ddot{\textmd{o}}$dinger equation to obtain
\begin{equation}
\begin{split}
\dot{g}_1(k)&=-\sqrt{\frac{\Gamma_1}{2\pi}}g_3e^{-it\delta_k},\\
\dot{g}_2(k)&=-\sqrt{\frac{\Gamma_2}{2\pi}}g_3e^{-it\delta'_k},\\
\dot{g}_3=&\int dk\left(\sqrt{\frac{\Gamma_1}{2\pi}}g_1e^{it\delta_k}+\sqrt{\frac{\Gamma_2}{2\pi}}g_2e^{it\delta'_k}\right),
\end{split}
\end{equation}
in which, $\delta_k=w_1-k$ and $\delta'_k=w_1-w_2-k$.

If the input photon has a Gaussian mode with bandwidth $d$ , the mode function can be described as
\begin{equation}
f(k)=(\frac{2}{\pi d^2})^{1/4}\exp[-\frac{(k-w_1)^2}{d^2}].
\end{equation}
As a result,  we can solve the Eq.(14) and obtain
\begin{equation}
g_2(k)=\int_0^tdt'\int_0^{t'}dt''-\frac{\sqrt{\Gamma_1\Gamma_2d}}{(2\pi)^{3/4}}e^{-it\delta'_k-\frac{d^2t''^2}{4}+\frac{\Gamma_1+\Gamma_2}{2}t''}.
\end{equation}

So the probability for generating the photon with central frequency $(w_1-w_2)$ is given by
\begin{equation}
\begin{split}
P(t)=&\int dk|g_2(k)|^2\\
=&\int_0^tdt'\frac{\Gamma_1\Gamma_2d}{(2\pi)^{1/2}}\left|\int_0^{t'}dt''e^{-\frac{d^2t''^2}{4}+\frac{\Gamma_1+\Gamma_2}{2}t''}\right|^2.
\end{split}
\end{equation}
\begin{figure}[h]
\includegraphics[scale=0.35]{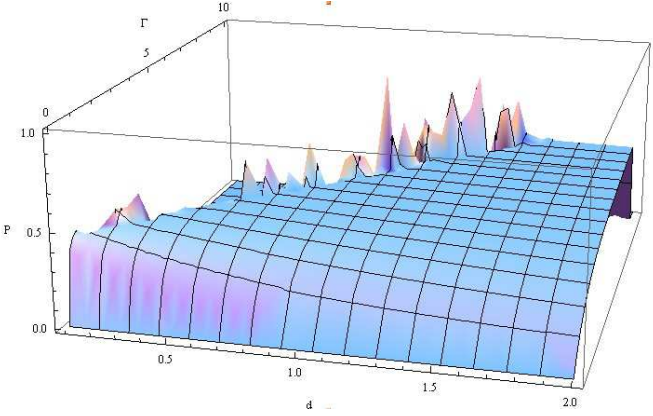}
 \caption{\label{fig.4} The numerical graph shows that the probability $P$ (for long time) changes with the bandwidth $d$ and the decay rate $\Gamma$ ($\Gamma=\Gamma_1=\Gamma_2$), where we use the  dimensionless unit.  }
 \end{figure}
From Fig. 4, we can find that the maximum probability $P$ is close to $1$. And one can input other photons to ensure the the maximum probability $P$ to be $1$. It means that swapping the multi-quantum-dot entanglement with multi-photon  can be with the high efficiency by this process.

Next, we try to create the polarization entanglement of multi-photon due to its advantage in the experiment.
\begin{figure}[h]
\includegraphics[scale=0.19]{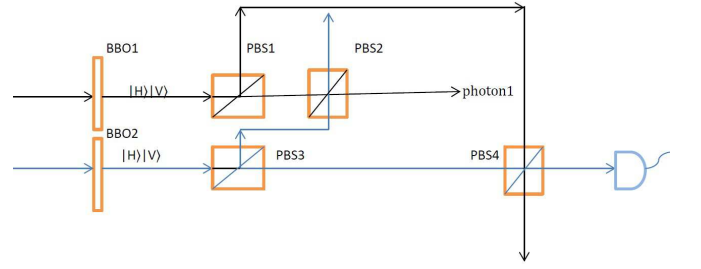}
 \caption{\label{fig.5} The input photons in the initial state $\frac{|10\rangle+|01\rangle}{\sqrt{2}}$ transmit through two $\beta$-barium borate (BBO) crystals, generating the state $\frac{|HV0\rangle+|0HV\rangle}{\sqrt{2}}$. Then, pass through four polarization beam splitters (PBS) to generate the polarization entanglement photon 1, which is ensured by the detection. Namely, if the detector finds a photon, then it means that the photon 1 is created. }
 \end{figure}
As shown in Fig. 5, the input photon state $\displaystyle\frac{|10\rangle+|01\rangle}{\sqrt{2}}$ can be converted into a polarization state $\displaystyle\frac{|H0\rangle+|0V\rangle}{\sqrt{2}}$ (photon 1). It can be treated as a single photon with polarization state $\displaystyle\frac{|H\rangle+|V\rangle}{\sqrt{2}}$. The small frequency difference between two polarization directions can be removed by the ultrafast downconversion technique \cite{lab12}.   In the way, the multi-photon entanglement $\displaystyle\frac{|10\rangle^{\otimes 2n}+|01\rangle^{\otimes 2n}}{\sqrt{2}}$ can be converted to the polarization state $\displaystyle\frac{|H\rangle^{\bigotimes n}+|V\rangle^{\bigotimes n}}{\sqrt{2}}$.

\section{conclusion}
 We have proposed a systematic scheme to create the multi-quantum-dot entanglement, and generate the multi-photon entanglement. Controlling the coupling between superconductor and quantum dots can effectively form the interaction for generating the entanglement, which is operable in the experiment. For the robust against the decoherence noise, the cavity modes are used to store the entanglement temporally. A error correction protocol can restore the coherence from the photons loss. It will assist the more qubits entanglement. Finally, use Gaussian photons to map the multi-quantum-dot entanglement into the multi-photon entanglement. And we find that the transform efficiency can be high with a few input photons.

 The scheme will help to create more photons entanglement beyond the present $8$ photons entanglement, which is useful in the quantum computation \cite{lab24}, simulation \cite{lab25,lab26}, and communication\cite{lab27}. This scheme can be performed with present-day technology. It stimulates the further study about the continuous-variable entanglement of photons, for example in Hilbert space of the angular momentum of photons \cite{lab28}. And multi-quantum-dot entanglement will also play an important role in the quantum computation.

\section*{ACKNOWLEDGMENT}
This work was supported by the National Natural Science Foundation of China under Grant No. 10975125 and No. 11375168.


\begin{thebibliography}{9}
\vspace{3mm}
\bibitem{lab1}Leggett and A. J., Rep. Prog. Phys. {\bf71}, 022001 (2008).
\bibitem{lab2}T. D. Ladd, F. Jelezko, R. Laflamme, Y. Nakamura, C. Monroe, and J. L. $\textmd{O}'$Brien,  Nature {\bf464}, 45-53 (2010).
\bibitem{lab3}Paul G. Kwiat, Klaus Mattle, Harald Weinfurter and Anton Zeilinger, Phys. Rev. Lett. {\bf75}, 4337-4341 (1995).
\bibitem{lab4}C. -Y. Lu, Xiao-Qi Zhou, Otfried G$\ddot{\textmd{u}}$hne, Wei-Bo Gao, Jin Zhang, Zhen-Sheng Yuan, Alexander Goebel, Tao Yang, and Jian-Wei Pan, Nature Phys. {\bf3}, 91-95 (2007).
\bibitem{lab5}R. Prevedel, G. Cronenberg, M. S. Tame, M. Paternostro, P. Walther, M. S. Kim, and A. Zeilinger, Phys. Rev. Lett. {\bf103}, 020503 (2009).
\bibitem{lab6} W. Wieczorek,  Roland Krischek, Nikolai Kiesel, Patrick Michelberger, G$\acute{\textmd{e}}$za T$\acute{\textmd{o}}$th, and Harald Weinfurter, Phys. Rev. Lett. {\bf103}, 020504 (2009).
\bibitem{lab7}M. Radmark, M. Zukowski, and M. Bourennane, Phys. Rev. Lett. {\bf103}, 150501 (2009).
\bibitem{lab8} J. C. F. Matthews, A. Politi, D. Bonneau and  J. L. O$'$Brien, Phys. Rev. Lett. {\bf107},
163602 (2011).
\bibitem{lab9}Yun-Feng Huang, Bi-Heng Liu, Liang Peng, Yu-Hu Li, Li Li, Chuan-Feng Li, and Guang-Can Guo, Nature Communications {\bf2}, 546 (2011).
\bibitem{lab10}Xing-Can Yao, Tian-Xiong Wang, Ping Xu, He Lu, Ge-Sheng Pan, Xiao-Hui Bao, Cheng-Zhi Peng, Chao-Yang Lu, Yu-Ao Chen and Jian-Wei Pan,  Nature Photonics {\bf6}, 225 (2012).
\bibitem{lab11}Thomas Monz,  Philipp Schindler, Julio T. Barreiro, Michael Chwalla, Daniel Nigg, William A. Coish, Maximilian Harlander, Wolfgang H$\ddot{\textmd{a}}$nse, Markus Hennrich, and Rainer Blatt, Phys. Rev. Lett. {\bf106}, 130506 (2011).
\bibitem{lab12}Kristiaan De Greve, Leo Yu, Peter L. McMahon, Jason S. Pelc, Chandra M. Natarajan, Na Young Kim, Eisuke Abe, Sebastian Maier, Christian Schneider, Martin Kamp, Sven H$\ddot{\textmd{o}}$fling, Robert H. Hadfield, Alfred Forchel, M. M. Fejer, and Yoshihisa Yamamoto, Nature {\bf491}, 421 (2012).
\bibitem{lab13}Martin Leijnse and Karsten Flensberg, Phys. Rev. Lett. {\bf111}, 060501 (2013).
\bibitem{lab14}Zaki Leghtas,  Gerhard Kirchmair, Brian Vlastakis, Robert J. Schoelkopf, Michel H. Devoret, and Mazyar Mirrahimi, Phys. Rev. Lett. {\bf111}, 120501 (2013).
\bibitem{lab15}Roman Kolesov,  Kangwei Xia, Rolf Reuter, Mohammad Jamali, Rainer St$\ddot{\textmd{o}}$hr, Tugrul Inal, Petr Siyushev, and J$\ddot{\textmd{o}}$rg Wrachtrup, Phys. Rev. Lett. {\bf111}, 120502 (2013).
\bibitem{lab16}Mahn-Soo Choi, C. Bruder and Daniel Los, Phys. Rev. A {\bf62}, 13569 (1999).
\bibitem{lab17}S. Nadj-Perge, S. M. Frolov, E. P. A. M. Bakkers, and L. P. Kouwenhoven, Nature (London) {\bf468}, 1084 (2010).
\bibitem{lab18}Y. Hu, F. Kuemmeth, C. M. Lieber, and C. M. Marcus, Nat. Nanotechnol. {\bf7}, 47 (2011).
\bibitem{lab19}Yong-Joo Doh, Jorden A. van Dam, Aarnoud L. Roest, Erik P. A. M. Bakkers, Leo P. Kouwenhoven,
 and Silvano De Franceschi, Science {\bf309}, 272 (2005).
\bibitem{lab20}Jorden A. van Dam, Yuli V. Nazarov,  Erik P. A. M. Bakkers,  Silvano De Franceschi, and  Leo P. Kouwenhoven, Nature (London) {\bf442}, 667 (2006).
\bibitem{lab21}Zaki Leghtas,  Gerhard Kirchmair, Brian Vlastakis, Michel H. Devoret, Robert J. Schoelkopf, and Mazyar Mirrahimi, Phys. Rev. A {\bf87}, 042315 (2013).
\bibitem{lab22}N. Aharon, M. Drewsen, and A. Retzker, arXiv:1307. 2933v1 (2013).
\bibitem{lab23}Kazuki Koshino, Phys. Rev. A {\bf79}, 013804 (2009).
\bibitem{lab24}R. Raussendorf, J. Harrington, and K. Goyal,  New J. Phys. {\bf9}, 199 (2007).
\bibitem{lab25}B. P. Lanyon, J. D. Whitfield, G. G. Gillett, M. E. Goggin, M. P. Almeida, I. Kassal, J. D. Biamonte, M. Mohseni, B. J. Powell, M. Barbieri, A. Aspuru-Guzik, and A. G. White, Nature Chem. {\bf2}, 106-111 (2010).
\bibitem{lab26}X. -S. Ma, B. Dakic, W. Naylor, A. Zeilinger, and P. Walther, Nature Phys. {\bf7},
399-405 (2011).
\bibitem{lab27}Ryszard Horodecki, Pawe${\l}$ Horodecki, Micha${\l}$  Horodecki, and Karol Horodecki, Rev. Mod. Phys. {\bf81}, 865 (2009).
\bibitem{lab28}Dong-Sheng Ding,	Zhi-Yuan Zhou,	Bao-Sen Shi, and Guang-Can Guo, Nat. Commun. {\bf4}, 2527	 (2013).
\end{thebibliography}
 \end{document}